\documentclass[a4paper,fleqn,usenatbib]{mn2e}
\usepackage{graphicx}
\usepackage{amsmath}
\usepackage{amssymb}
\usepackage{upgreek}

\title[Pulsar braking and the $P-\dot{P}$ diagram]
{Pulsar braking and the $P-\dot{P}$ diagram}

\author[Johnston \& Karastergiou]  {Simon Johnston$^{1,2}$\thanks{email: Simon.Johnston@csiro.au} and A. Karastergiou$^{3,4,5}$
\\
$^{1}$CSIRO Astronomy and Space Science, Australia Telescope National Facility, PO Box 76, Epping, NSW 1710, Australia\\
$^{2}$Max-Planck-Institut f\"ur Radioastronomie (MPIfR), Auf dem H\"ugel 69, D-53121 Bonn, Germany\\
$^{3}$Oxford Astrophysics, Denys Wilkinson Building, Keble Road, Oxford, OX1 3RH, UK.\\
$^{4}$Physics Department, University of the Western Cape, Cape Town 7535, South Africa\\ 
$^{5}$Department of Physics and Electronics, Rhodes University, PO Box 94, Grahamstown 6140, South Africa
}
\date{Accepted \today. Received \today; in original form \today}

\pubyear{2016}

\begin{document}
\label{firstpage}
\pagerange{\pageref{firstpage}--\pageref{lastpage}} 
\maketitle

\begin{abstract}
The location of radio pulsars in the period-period derivative ($P-\dot{P}$)
plane has been a key diagnostic tool since the early days of pulsar
astronomy. Of particular importance is how pulsars evolve through
the $P-\dot{P}$ diagram with time. Here we show that the decay of the
inclination angle ($\dot{\alpha}$) between the magnetic and rotation axes plays 
a critical role. In particular, $\dot{\alpha}$ strongly impacts on the
braking torque, an effect which has been largely ignored in previous work.
We carry out simulations which include a negative $\dot{\alpha}$ term,
and show that it is possible to reproduce the observational $P-\dot{P}$ 
diagram without the need for either pulsars with long birth periods or
magnetic field decay.  Our best model indicates a 
birth rate of 1 radio pulsar per century and a total Galactic population of 
$\sim$20000 pulsars beaming towards Earth.
\end{abstract}


\section{Introduction}
\begin{figure}
\includegraphics[width=8cm]{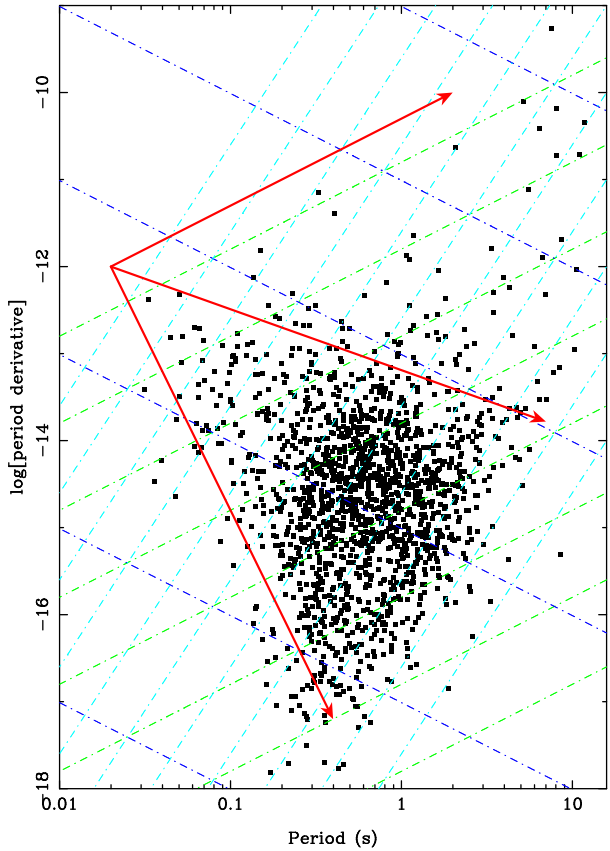}
\caption{The $P-\dot{P}$ diagram for 1600 known pulsars. Lines of constant
$B$ are in blue, line of constant $\tau_c$ are green and line of constant
$\dot{E}$ are light blue. From an initial position at $P=20$~ms, 
$\dot{P}=10^{-12}$, the red arrows show time-evolution
through the diagram for $n=1.0$, 2.7 and 6.0 from top to bottom.}
\label{ppdot}
\end{figure}

Upon the discovery of a radio pulsar, its position in the sky,
its spin period, $P$, and dispersion measure are immediately known. The 
technique of pulsar timing subsequently allows the slow-down rate, $\dot{P}$,
to be determined. From the very early days of pulsar astronomy, therefore,
pulsars could be placed on the $P-\dot{P}$ plane.  Figure~\ref{ppdot} shows 
the modern $P-\dot{P}$ diagram for 1600 of the known pulsars. In this figure
we have excluded all re-cycled pulsars to concentrate on the bulk
of the slow pulsar population.
\begin{figure}
\includegraphics[width=8cm]{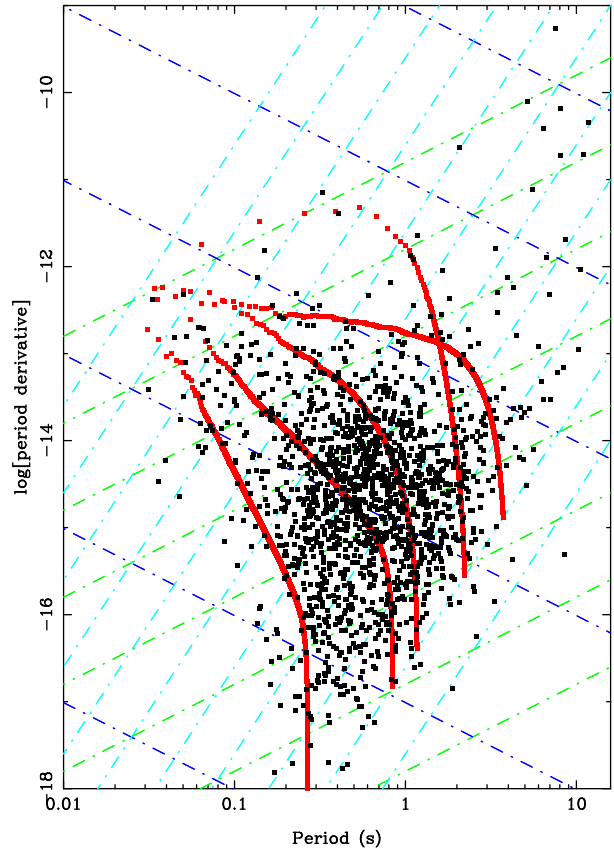}
\caption{As for Figure~\ref{ppdot}.  From an initial
position at $P=20$~ms, $\dot{P}=10^{-12}$, the red lines denote time-evolution
through the diagram in steps of 1000~yr according to Equation~\ref{ttraw} for 
different initial values of $n$. Evolutionary tracks end once the pulsar
has crossed the death line.}
\label{tracks}
\end{figure}

If we assume that the pulsar is a magnetic dipole rotating in a vacuum,
the surface magnetic field strength, $B$, is given by
\begin{equation}
B = \sqrt{\frac{3c^3I}{8\pi^2R^6{\rm sin}^2\alpha} P \dot{P}}
\label{bfield}
\end{equation}
where $c$ is the speed of light, $I$ is the moment of inertia of the star,
$R$ is its radius and $\alpha$ is the inclination angle between the 
rotation and magnetic axes.
If therefore one assumes that $I$ and $R$ are the same for all pulsars
and ${\rm sin}^2\alpha = 1$, one can draw lines of constant $B$
onto the $P-\dot{P}$ diagram (see Figure~\ref{ppdot}).
Similarly, the spin-down energy, $\dot{E}$, can be written
\begin{equation}
\dot{E} = 4\pi^2 I \frac{\dot{P}}{P^3}
\end{equation}
and again this allows for lines of constant $\dot{E}$ on the $P-\dot{P}$ 
diagram. Finally, the characteristic age, $\tau_c$ of the pulsar is computed via
\begin{equation}
\tau_c = \frac{P}{2\dot{P}}
\end{equation}
The value of $\tau_c$ is equal to the true age under the assumption that the
initial spin period of the pulsar is much less than its current period
and that dipolar magnetic braking is the sole cause of the spin-down.
Lines of constant $\tau_c$ are also included in Figure~\ref{ppdot}.

Under the assumptions made above, the $P-\dot{P}$ diagram can be used
as an evolutionary tool. Young pulsars live at the top left of the diagram
with small $P$ and high $\dot{P}$. The magnetars, with their high $B$-fields
live in the top right of the diagram. The bulk of the pulsars form a 
roughly circular shape in the diagram. Very few pulsars have $\dot{E}$
below $10^{30}$~ergs$^{-1}$; this marks the so-called death-line below which
it is believed radio emission ceases to be viable.

More generally, the spin-down of a pulsar can be written in the form
\begin{equation}
\dot{\nu} = -K \nu^n
\label{nu}
\end{equation}
where here $\nu$ and $\dot{\nu}$ are the spin frequency and its derivative,
$K$ is constant and $n$ is the braking index.
Taking the time derivative of Equation~\ref{nu} yields
\begin{equation}
n = \frac{\nu \ddot{\nu}}{\dot{\nu}^2}
\label{brake}
\end{equation}
and so $n$ can in principle be measured if $\ddot{\nu}$ can be obtained.
If both $K$ and $n$ are constant in time, a pulsar will then follow
a track in the $P-\dot{P}$ diagram with a slope of $2-n$.
Theoretical expectations are that if the torque is dominated by an
outflowing wind then $n=1$, if magnetic dipole dominated then 
$n=3$, and if magnetic quadropole dominated then $n=5$
in the absence of other effects \citep{ac04}.
\citet{hsuu15} have shown that $n$ can deviate from these values
because of the changing $I$ as the star spins down.

Two important modifications to this simple picture, magnetic field decay
and alignment of the spin and magnetic axes, were outlined in \citet{tk01}. 
They showed that, in this case, the braking index is a function of time and 
depends on the time-evolution of $B$ and $\alpha$ in the following way:
\begin{equation}
n(t) = 3.0 - \frac{3 c^3 I \dot{B}(t)}{R^6 B^3(t) {\rm sin}^2\alpha(t) \Omega^2(t)} - \frac{3 c^3 I {\rm cos}\alpha(t) \dot{\alpha}(t)}{R^6 B^2(t) {\rm sin}^3\alpha(t) \Omega^2(t)}
\label{ttraw}
\end{equation}
In this equation, the second term contains the time derivative of the
magnetic field, $\dot{B}$, with the third term relating to the time
derivative of the inclination axis, $\dot{\alpha}$.

Unfortunately $\ddot{\nu}$ is small and difficult to measure, making $n$ hard
to determine with any accuracy in all but a handful of young pulsars.
Measured values of $n$ range from $0.9$ in
PSR~J1734--3333 \citep{elk+11} to $3.15$ for PSR~J1640--4631 \citep{agf+16}
though \citet{mgh+16} have recently reported $n$ close to zero for
PSR~B0540--69 in the Large Magellanic Cloud.
Astonishingly, the braking index of PSR~J1846--0258 changed from
$2.65$ to $2.19$ in less than a decade \citep{akb+15} and a smaller
change was seen in PSR~J1119--6127 \citep{awe+15}. This implies that
substantial torque changes can be applied to the star on short timescales.
We also note that \citet{jg99} proposed a way to obtain $n$ without the
need to measure $\ddot{\nu}$ in a fully coherent solution.
In their paper, they reported a number
of pulsars with high values of $|n|$ (and small error bars)
over the timescale of a decade.
They surmised that these high values were caused by recovery from (unseen) 
glitches but whatever the cause, high values of $n$ are clearly plausible.

In addition to direct measurements of $n$, there is a mounting body of
evidence for torque changes on short timescales in many, if not all, pulsars.
The class of pulsars known as `intermittents' show that $\dot{P}$ changes
significantly between the `on' and `off' states likely due to the presence
of plasma in the magnetosphere \citep{klo+06}. \citet{lhk+10} showed
state (profile) changes accompanied by $\dot{P}$ changes in a large number 
of pulsars, a study backed up by \citet{bkj+16}. Torque changes may also
be induced through interaction with asteroids \citep{scm+13,bkb+14}
or free precession \citep{khjs16}.
With these results in mind, we therefore modify Equation~\ref{ttraw} to read
\begin{equation}
\begin{aligned}
n(t) = {} & n_0 - \frac{3 c^3 I \dot{B}(t)}{R^6 B^3(t) {\rm sin}^2\alpha(t) \Omega^2(t)} \\
 & -\frac{3 c^3 I {\rm cos}\alpha(t) \dot{\alpha}(t)}{R^6 B^2(t) {\rm sin}^3\alpha(t) \Omega^2(t)} + {\rm TN}(t)
\label{tt}
\end{aligned}
\end{equation}
where now $n_0$ is the initial braking index and ${\rm TN}(t)$ is a random
component of the braking index due to effects of state changes, timing
noise and/or intermittency. This will be discussed further in Section~3.

A major unresolved challenge in pulsar astronomy is determining where pulsars 
are born in the $P-\dot{P}$ diagram and how then they evolve through the
diagram with time. We explore this issue further in Section~2. In
Section~3 we outline our simulation, present the results in Section~4,
discuss the implications in Section 5 before concluding in Section~6.

\section{Evolution in the $P-\dot{P}$ diagram}
Two major problems confront us when we study the $P-\dot{P}$ diagram.
The first is that young pulsars such as the Crab pulsar
($P\simeq 33$~ms, $\dot{P}\simeq 4\times10^{-13}$) appear to have
magnetic fields larger than older pulsars. Hence, one idea is that the 
magnetic field decays as pulsars get older as was first 
postulated by \citet{go70}.
This remains controversial as more modern studies still press the case for
(e.g. \citealt{gvh04}) and against (e.g. \citealt{lbh97}) field decay.
Theoretical work in this area also has not reached a consensus,
suggesting field decay is either unimportant
\citep{gr92} or relatively rapid \citep{gmv14,ip15} over the pulsar lifetime.
We note that field decay almost certainly occurs during episodes of
mass accretion but this is not relevant to the general population of
isolated pulsars under consideration here.
The second issue is whether all pulsars are born
like the Crab pulsar with a fast ($\simeq$20~ms) initial spin period and high
magnetic field, or whether
the $P-\dot{P}$ diagram can only be explained by postulating an
`injection' of pulsars with much slower ($\sim$500~ms) birth periods.
This idea, first championed by \citet{vn81}, still has recent proponents
(e.g. \citealt{vml+04}) but other studies find no need for long
period at birth (e.g. \citealt{gob+02}).

Population studies which attempt to replicate the properties of the
observed pulsars follow two different approaches. The first is to
take a snapshot of the Galaxy as it appears today and match it to
the observed popuation (see e.g. \citealt{blr+14}). The second is an
{\it ab initio} approach which has evolution of pulsars from birth onwards.
As a fine example of the latter genre, we consider the comprehensive study of 
the birth and evolution of isolated radio pulsars by \citet{fk06}.
The major conclusions of their paper are that (i) magnetic field decay
is not significant, (ii) the luminosity $L$ of a pulsar is proportional
to $\sqrt{\dot{E}}$, (iii) the initial spin period has a mean of 300~ms
with a wide distribution.
One of the assumptions in \citet{fk06} is that the braking index of
a pulsar has a constant value over the pulsar's lifetime and although
they explored a distribution in this constant they did not attempt
to model evolution of $n$ as a function of time.
Similarly, more recent work by \citet{rl10} investigated random values
for $n$ at birth but did not consider $n$ to vary with time.
Both groups include $\alpha$ decay in their simulations,
and yet failed to include time-variable $n$ even though this is implied
via Equation~\ref{tt}.

A more theoretical approach was taken by \citet{gmv14} and they take
proper account of the time evolution of the magnetic field and the inclination
angle. In addition, they include the effects of plasma in the magnetosphere
and further consider the magneto-thermal evolution of the star. They
strongly favour a model where $\dot{B}<0$ and a power-law decay of $\alpha$.

\subsection{Evidence for non-zero $\dot{\alpha}$}
Only one direct observational measurement of $\dot{\alpha}$ has been made
in a normal, isolated pulsar.
\citet{lgw+13} reported on 22 years of timing of the Crab 
pulsar which shows that $\alpha$ is increasing at a rate of 0.62 degrees
per century. Statistically, however, it appears the opposite is the case.
\citet{tm98} examined the distribution of $\alpha$ in the known pulsar
population under the assumption of filled circular beams and measurements
of the pulse width. They showed that there were many more small values
of $\alpha$ than expected and concluded that alignment must occur (i.e.
that $\dot{\alpha} < 0$) on a timescale of $10^7$~yr.
\citet{wj08} came to very similar conclusions by attacking the problem
in a different way. They showed that the fraction of pulsars with
interpulses was inconsistent with a random distribution of $\alpha$. They
concluded that although $\alpha$ was random at birth, $\dot{\alpha}$ must
be less than zero so that the magnetic and rotational axes align with time.
\citet{jon76} argues that alignment only begins to occur when the pulsar
age exceeds $10^4$~yr, once the temperature-dependent dissipative torque
becomes negligible. If this is correct, the expectation is that the value of
$\dot{\alpha}$ in the Crab will change sign in the future.
The form of $\alpha$ decay has been considered theoretically by
\citet{ptl14} for pulsars with and without plasma loading of the
magnetosphere. For the vacuum case, pulsars align (unrealistically) fast.
With plasma loading the alignment time is much longer and the
form of the decay is power-law rather than exponential. Even in this case,
however, the timescale is shorter than observers are comfortable with.
In light of the uncertainties in the theoretical models, we
assume an exponential decay of $\alpha$ with a timescale of order $10^7$~yr.

We note that in the case of the very old millisecond pulsars and the
neutron stars in X-ray binary systems, that $\alpha \neq 0$.
However, these pulsars have had episodes of mass accretion from their
companion star. Mass accretion is expected to cause magnetic field
decay in these systems (as first postulated by \citealt{acrs82}) and almost 
certainly causes a re-arrangement of the mangetic field structure and 
hence $\alpha$ \citep{pm04}.

\section{Simulations}
The thinking behind the introduction of a variable braking index is as follows:
if all pulsars are born with parameters similar
to that of the Crab pulsar, can we use braking index alone to replicate
the $P-\dot{P}$ diagram without the need for either magnetic field decay or
for pulsars born spinning slowly?
We have seen that since the publication of \citet{fk06}, our knowledge
of the braking indices of pulsars has changed dramatically. It now appears
evident that not only is there a wide range of $n$ but that $n$ can
change significantly even on short timescales. Furthermore, as listed above,
the evidence for $\alpha$-decay is strong and this implies a long-term
evolution of $n$ as shown by \citet{tk01}.

First consider straight line tracks in the $P-\dot{P}$ diagram (recalling that 
for a constant $n$, the evolution in the diagram follows a slope of $2-n$)
for a pulsar with initial parameters of $P=20$~ms, $\dot{P}=10^{-12}$.
In order for this pulsar to reach the bottom left of the bulk of the population,
$n$ must be 6.0.  For it to reach the bottom right of the population,
$n$ must be 2.7 while to reach the magnetars at the extreme top right then
$n$ must be 1.0. Figure~\ref{ppdot} shows these tracks. Now consider
that $n$ can vary with time as in Equation~\ref{ttraw}. Figure~\ref{tracks}
shows five example tracks with $\dot{B}=0$ and an exponential decay of
$\alpha$ on a timescale of $10^7$~yr.

Several ideas then manifest themselves. The first is that $n$
is fixed at birth, but takes a wide range of values sufficient to
be able to populate the $P-\dot{P}$ plane. This idea was tested by
\citet{fk06} and \citet{rl10} and found not to provide a good match to 
the observed population. The second is that $n$
be time dependent and perform a random walk over the allowed
parameter space. A simple simulation therefore involves using Equation~\ref{tt}
with $\dot{B}=\dot{\alpha}=0$, picking an
initial value for $n_0$ and then allowing ${\rm TN}(t)$ to provide short
timescale variations.
Finally, $n$ varies according to Equation~\ref{tt} and different 
combinations of $\dot{B}$ and $\dot{\alpha}$ can be trialled in addition
to the random component ${\rm TN}(t)$.

In our simulation, we build on the work of others by fixing many of the
initial parameters. For the spatial distribution of pulsars we assume 
the form of the radial distribution given by \citet{lor04} 
\begin{equation}
\rho_r(R) = K_r \,\,\,  R^i \,\,\, e^{-R/\sigma_r}
\end{equation}
where $\rho_r(R)$ is the density of pulsars (per kpc$^2$)
at radius $R$ (in kpc) from the 
Galactic Centre and $K_r$, $i$ and $\sigma_r$ are constants with
values of 64.6~kpc$^{-2}$, 2.35 and 1.258~kpc respectively.
For the $z$-height distribution we use an exponential 
with a scale height of 330~pc \citep{lfl+06}. For any given pulsar, therefore,
we pick a random distance from Earth, $d$, based on these distributions.
We assume that $\alpha$ is randomly distributed at birth 
(i.e. that the probability distribution is sin($\alpha$), see also
\citealt{gmv14}) but exponentially
decays towards $\alpha = 0$ with a time constant of 
$5\times10^7$~y \citep{tm98,wj08}.  This is a crucial feature of our model 
because the beaming fraction of pulsars with low $\alpha$ 
is smaller than at high $\alpha$ and we have seen from Equation~\ref{tt}
how $\dot{\alpha}$ affects the value of $n$.
We take the half-opening angle $\rho$ of a pulsar's radio beam
(in degrees) at a canonical observing frequency of 1.4~GHz to be
\begin{equation}
\rho = 6.8 P^{-0.5}.
\label{rho}
\end{equation}
This is slightly larger
than generally assumed \citep{kwj+94} but in line with the results
of \citet{mr02}. The combination of $\alpha$ and $\rho$ yields a
beaming fraction, the probability that the pulsar is beaming towards Earth
(see e.g. \citealt{tm98}). If the pulsar is beaming towards Earth
we then pick a value of the impact parameter, $\beta$ between $-\rho$
and $+\rho$. The combination of $\rho$, $\alpha$ and $\beta$ then
yields the observed pulse width, $w$ (e.g. \citealt{ggr84}).
We also introduce a death line at $\dot{E} = 10^{30}$~ergs$^{-1}$ as do
\citet{fk06}.

\begin{figure}
\includegraphics[width=8cm]{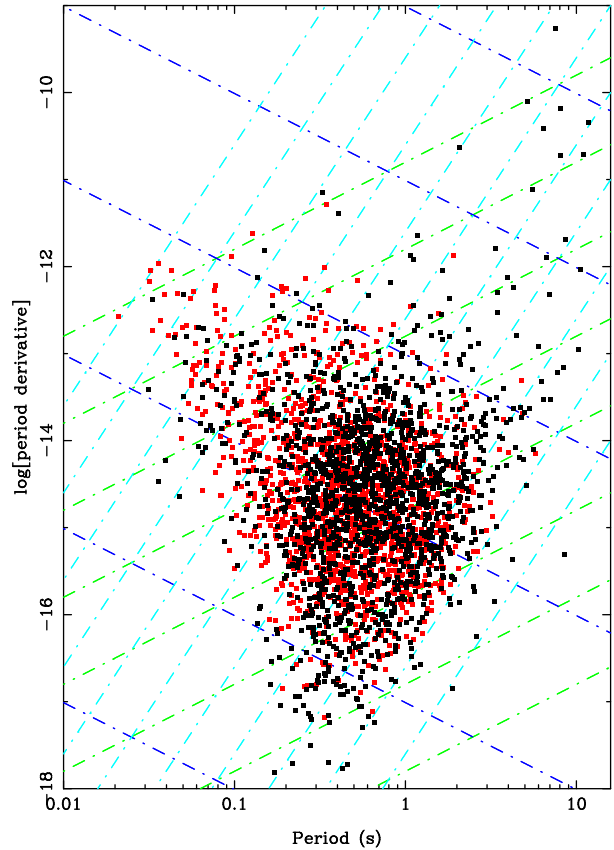}
\caption{$P-\dot{P}$ diagram for the known pulsars (black points) and
the pulsars detected in the simulation (red points).}
\label{compare}
\end{figure}
We want to test the idea that all pulsars are born like the Crab, so we fix 
the birth parameters at $P=20$~ms, $\dot{P}=10^{-12}$.
Pulsar ages are evenly distributed in steps of the birth 
rate, $B_R$, up to some maximum age, $T_{\rm max}$ so that the number of 
pulsars is $T_{\rm max}/B_R$. In principle $T_{\rm max}$ could be set
to the age of the Galaxy; in practice we find that $10^8$~y is sufficient as 
pulsars older than this are either too faint to be detectable or fall below
the death line. The value of $B_R$ ranges from 3 per century to less than
1 per century with the arguments summarised in \citet{kk08}. We return
to the value of this parameter in the next section.

This leaves us requiring a luminosity law, in order to determine whether
or not a pulsar beaming towards Earth is actually detectable giving the 
sensitivity of radio telescopes. Discussion over the form of the 
luminosity distribution
of radio pulsars has continued in the literature for more than 30 years,
and is well summarised in \citet{fk06}. An accepted form for the
luminosity law is
\begin{equation}
{\rm log}L = {\rm log}(L_0 \,\,\, P^{\epsilon_1} \,\,\, \dot{P}^{\epsilon_2})   + L_c.
\label{lumin}
\end{equation}
In the recent literature,
\citet{fk06} and \citet{gmv14} have $\epsilon_1 = -1.5$, $\epsilon_2 = 0.5$
whereas \citet{rl10} have $\epsilon_1 = -1.0$, $\epsilon_2 = 0.5$ and
\citet{blr+14} have $\epsilon_1 = -1.4$, $\epsilon_2 = 0.5$.
Once the luminosity is known we can convert to a (pseudo-) flux density;
$S = L d^{-2}$.

It is beyond the scope of this paper to reproduce the complex selection 
effects of real pulsar surveys. Rather, we make the conversion from 
luminosity to signal-to-noise ratio (SNR) via
\begin{equation}
{\rm SNR} = \frac{L}{d^2 S_0} \sqrt{\frac{P-w}{w}}
\label{snr}
\end{equation}
using a simple scaling term $S_0$ which is adequate for our purposes.
The term inside the square root comes from 
the fact that pulsars are not found via continuum imaging, but rather through
a Fourier technique which has the consequence that narrow pulses are easier 
to detect than broad pulses of the same power \citep{dtws85}.

The simulation therefore proceeds as follows. Initial parameters
$P$, $\dot{P}$, $d$ and $\alpha$ are chosen. The initial braking index, $n_0$
is drawn from a normal distribution with a mean of 2.8 and a $\sigma$ of 1.0.
The pulsar evolves in time, and values of $P$, $\dot{P}$, $\alpha$
and $n$ (according to Equation~\ref{tt}) are updated. The form of ${\rm TN}(t)$
is such that every 1000~y, a random value is picked from a Gaussian
distribution with a mean of the current value of $n$ and a $\sigma$ of
$n/3$.
Once the pulsar has reached the appropriate age, $\rho$, $\beta$, $\dot{E}$,
$L$ and SNR are computed.
A pulsar is deemed detectable if (a) it is beaming towards 
Earth, (b) it has ${\rm SNR}>10$ (according to Equation~\ref{snr} above)
and (c) its $\dot{E}$ places it above the death line. 
The ensemble of detected pulsars can then be compared with the known
pulsar population.

\section{Results of the simulation}
\subsection{Luminosity law and birth rate}
As with all population studies, the parameters of the luminosity law
(Equation~\ref{lumin}) are critical to the output of the simulation.
We set $L_c=0$. Clearly, in the observed population there can be large
differences in luminosity for pulsars with similar $P$ and $\dot{P}$. In a 
statistical sense however, $L_c$ makes little difference to the outcome
of the simulations.

We initially tested $\epsilon_1 = \epsilon_2 = 0.0$ so that the
luminosity is independent of $P$ and $\dot{P}$. This yielded a
large number of detected pulsars with long periods and a `pile-up' of
detections close to the death line.
This is clearly in disagreement with the observations,
a conclusion also reached by others \citep{fk06,rl10}.

We then tested $\epsilon_1 = -1.5$, $\epsilon_2 = 0.5$, the values
favoured by \citet{fk06}. This is appealing
because it implies that $L\propto\dot{E}^{1/2}$ similar to that seen
in $\gamma$-ray pulsars \citep{pmc+13}. We find that this luminosity law
results in the detection of too many young, short period pulsars compared
to the observed distribution. This is because $\dot{E}$ decreases by 8 orders
of magnitude over the lifetime of a pulsar and so the luminosity law
is strongly biased towards high $\dot{E}$ pulsars. To counteract this problem,
the general solution is to postulate long periods at birth as both
\citet{fk06} and \citet{blr+14} do. 

Clearly though if $\epsilon_1 = \epsilon_2 = 0.0$ produces too many old
pulsars and $\epsilon_1 = -1.5$, $\epsilon_2 = 0.5$ produces too many
young pulsars, an alternative solution would be to have a luminosity
law somewhere in between these two possibilities.  Indeed, we find an 
acceptable fit to the observed pulsars can be made by setting
$\epsilon_1 = -0.75$, $\epsilon_2 = 0.25$. 

We find the best fit to the data is obtained with $B_R$ set to 1 per century.
Higher values of $B_R$ result in the detection of too many short period
pulsars compared to the observed population.
Our value is significantly lower than the 2.8 per century found by \citet{fk06}
and the $\sim$3 per century in \citet{gmv14}
but within the errors of the \citet{lfl+06} and \citet{vml+04} values.

\subsection{Model performance}
Figure~\ref{compare} shows the comparison between the simulation and
the known pulsar population.  As \citet{fk06} discussed, it is
difficult to quantitively assess the goodness-of-fit for these type of
simulations. The Kolmogorov-Smirnov (K-S) test is rather a blunt
instrument and does not work well for multi-dimensional data. Although
the K-S test could be used on individual parameters such as $P$ and
$\dot{P}$, these parameters are not independent making it hard to
judge the output from the K-S test. We therefore proceeded as follows:
We generate 2D histograms in the $P-\dot{P}$ plane for both the
observed and simulated populations. The histograms are generated
using a set of 20 bins of equal width in logarithm space along each
direction, and counting the number of simulated pulsars
($N_{\rm sim}$) and the number of observed pulsars ($N_{\rm obs}$) for
each 2D bin.  We then define the significance of the difference in the
pairs of numbers, $R$, through
\begin{equation}
\label{eqnS}
R = \frac{N_{\rm sim} - N_{\rm obs}}{\sqrt{N_{\rm sim} + N_{\rm obs}}}
\end{equation}
and assign a colour scale to the range of values of $R$. 
\begin{figure}
\includegraphics[width=8cm]{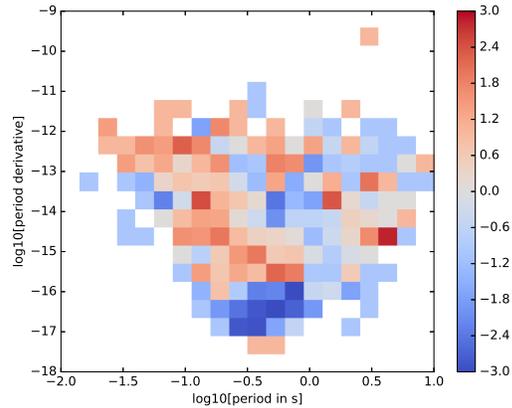}
\caption{The difference between the observed and simulated $P-\dot{P}$
diagram, colour coded using Eq.~\ref{eqnS}. Red signifies an over-abundance
of simulated pulsars, blue an under-abundance.} 
\label{heatmap}
\end{figure}

Figure~\ref{heatmap} shows a visual representation of equation~\ref{eqnS}
and although the figure cannot be used in a statistical sense, it
is indicative of the goodness-of-fit of the model.
The model performs most poorly
towards the bottom left of the $P-{\dot P}$ diagram, where the
simulation underpredicts the observed numbers.  \citet{lma+04}
speculate that this part of the diagram contains pulsars originally
part of a binary system. Their magnetic field then decays and their
spin period decreases as a result of accretion of material from the
binary companion before the system disrupts after the second supernova
\citep{blrc10}.  This provides a plausible explanation for the surfeit
of pulsars in this part of the diagram.  Our model also somewhat
underestimates the magnetar population, those pulsars with long
periods and high $\dot{P}$. Although some evolutionary tracks head in
the right direction (see Figure~\ref{tracks}), $\alpha$ decay pulls
them downwards before they reach the magnetar area of the $P-\dot{P}$
diagram. It remains unclear whether magnetars are a separate class of
pulsars or whether indeed they arise from evolution from a different
part of $P-\dot{P}$ space \citep{kk08,elk+11}. Furthermore, magnetars
appear to have wide beams and do not conform to
Equation~\ref{rho}. This presumably implies that their beaming
fraction is much larger than our simulation supposed, causing our
model to underestimate their population.

\section{Discussion}
\subsection{Differences with other studies}
There is strong evidence that $\alpha$ decays with time
\citep{tm98,wj08,ptl14} and therefore that $n$ is time-variable according to 
Equation~\ref{tt}. This is an integral part of our model.
\citet{rl10} consider $\alpha$ decay in detail but they
do not make the important connection between this and the braking index.
Their pulsars therefore do not move correctly in $P-\dot{P}$ space.

The optimum model of \citet{fk06} does not include $\alpha$ decay and
assumes a constant braking index of 3. Pulsars are therefore forced
to move along lines of constant $B$ (see Figure~\ref{ppdot}).
The observed pulsars have a wide
range of $B$ and so their model must reflect this in the birth parameters.
Indeed their model has $\sigma=0.55$ (in the log) for the birth $B$ field.
In addition, their luminosity law forces pulsars to be born with relatively
long initial spin periods to prevent an overabundance of short period
pulsars. In our model, pulsars have high $B$ and short $P$ at birth.
Although $B$ does not decay, the {\it apparent} $B$ (as computed via
Equation~\ref{bfield}) drops because $n$ becomes larger than 3 as the pulsar
ages. At the same time, we have flattened the dependence of $L$ on $P$,
which alleviates the young pulsar problem. It does not create an old pulsar
problem, because old pulsars have smaller $\alpha$ which reduces their
beaming fraction and increases the pulse width which reduces their
detectability.

The work of \citet{gvh04} includes the effect of $B$-field decay with
their optimal model having a decay timescale of only 2.8~Myr. Initial
spin periods are short, and initial values of $B$ are generally higher
than found by other groups. They also include a full description of core-cone 
radio beams which modifies their luminosity law which has
$\epsilon_1 = -1.3$, $\epsilon_2 = 0.5$. They do not consider $\alpha$ decay.
In some ways, our results reflect theirs; they set $\dot{\alpha}=0$ and
have $\dot{B}<0$ whereas we have $\dot{\alpha}<0$ and
$\dot{B}=0$ which has the same effect on $n$ according to Equation~\ref{tt}.
However, as explained above smaller values of $\alpha$ for older pulsars
reduces their detectability which we believe is a crucial difference
in the modelling.

\begin{figure}
\includegraphics[width=6cm,angle=-90]{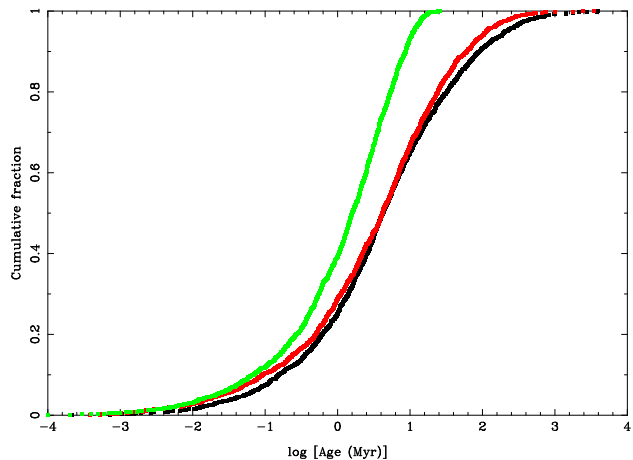}
\caption{Cumulative distribution of the characteristic ages of the
known population (black) and the simulated detections (red) with the
true ages of the simulated pulsars shown in green.}
\label{age}
\end{figure}

Finally \citet{gmv14} include both $B$-field and $\alpha$ decay in their
models. Although their results are rather agnostic as to $B$-field decay
they strongly prefer a power-law decay of $\alpha$. They have a similar 
luminosity law to \citet{fk06} and hence a large number of pulsars with 
long initial periods. We disagree with their findings that short initial
periods cannot reproduce the $P-\dot{P}$ diagram.

\subsection{Implications}
The idea that a time-variable braking index is a key component in the
evolution of pulsars in the $P-\dot{P}$ diagram leads to some testable
predictions. First, as seen in Figure~\ref{tracks}, old pulsars move
vertically in $P-\dot{P}$ space and should have large values of $n$.
For a pulsar with $P\sim ~1$s, $\dot{P}\sim 10^{-15.5}$ and $n\sim 1000$
it may be possible to measure $\ddot{\nu}$ over a 20~yr period.
Intriguingly, in the compilation of \citet{jg99}, the two pulsars with the 
largest values of $P$ also have large values of $n$.
Secondly, pulsars with smaller values of $\alpha$ will have larger
values of $n$ and this also applies to young pulsars where $n$ is more
easily measurable. A further interesting test would therefore be to
try and measure $n$ for a group of young pulsars with known $\alpha$
such as the sample of \citet{rwj15}.

In our simulation, all pulsars are born with $P=20$~ms. Clearly this is a 
simplification of the true picture but the observational evidence is strong
that initial spin periods are less than 150~ms for a large fraction of
the population. In contrast to our results,
the best model of \citet{fk06} has 84\% of their pulsars have spin periods
larger than 150~ms and similarly \citet{gmv14} have a uniform distrubution
of initial periods between 0 and 150~ms as their best fit.
In both papers this appears to be a direct result of their luminosity law and
it is hard to reconcile this high fraction with the observational evidence
from the known young pulsars.

The results also have implications for the lifespan of pulsars and hence
the potentially detectable population. In \citet{fk06}, $\tau_c$ is
equivalent to the true age and many pulsars live well in excess of $10^8$~yr
before crossing the death line. This is not the case in our work; in
general $\tau_c$ is significantly greater than the true age, especially
for older pulsars. For example, only 7\% of the detected population
from the simulation have true ages greater than $10^7$~yr whereas
35\% have $\tau_c > 10^7$~yr, similar to that of the known population.
Figure~\ref{age} shows the cumulative distribution of the characteristic
and true ages of the pulsars.
In addition to the age differences, $\alpha$ decay reduces the
number of pulsars beaming towards Earth. Our modelling therefore suggests
that the population of radio-loud pulsars is only $2\times 10^5$ and
only 10\% of these are beaming towards us, a factor of $\sim$5 less than
the \citet{fk06} result but close to the \citet{lfl+06} result.
This indicates that estimates for how many pulsars the Square Kilometre
Array will find \citep{kbk+15} may be somewhat overestimated.

\section{Conclusions}
Nearly fifty years after the discovery of the first radio pulsar, their
time-evolution remains a source of debate. Two topics in particular
remain a constant thread, the first being the decay (or not) of the magnetic
field, the second being their birth parameters. The literature generally
favours a lack of field decay except through accretion and relatively
large spin periods at birth making the Crab pulsar an exception.

In this paper we show the importance of the decay of the inclination
angle between the magnetic and spin axes, an idea which has firm footing in 
the literature. This decay has two main effects. First it reduces the
detectability of older pulsars, as small $\alpha$ reduces the beaming
fraction and increases the observed pulse width. Secondly, it modifies
the spin-down evolution so that pulsars no longer follow a track with
constant slope in the $P-\dot{P}$ diagram.

Although we have not attempted a full-blown population analysis, we have
shown that it is possible to have all pulsars be born with parameters
similar to that of the Crab pulsar and still reproduce the bulk of the 
features of the known population obviating the need for magnetic field
decay or pulsars with long initial spin periods.
If these ideas are correct, the birth rate of pulsars is 1 per 100~yr and
there are only $\sim$20000 pulsars beaming towards Earth.

\section*{Acknowledgments}
We thank M. Kramer and T. Tauris for useful discussions. We used the
ATNF pulsar catalogue at http://www.atnf.csiro.au/people/pulsar/psrcat/
for this work.

\bibliographystyle{mnras}
\bibliography{ppdot}
\bsp
\label{lastpage}
\end{document}